\documentclass{blois}

\usepackage{lipsum}
\usepackage{mathrsfs}
\usepackage{soul}
\usepackage{amsmath,amssymb,mathtools,bm}
\usepackage{graphicx, color, hepunits}
\usepackage[dvipsnames]{xcolor}
\usepackage{float}
\usepackage{multirow}
 \usepackage{hyperref}

\def\bk{\boldsymbol{k}}

\def\bx{\boldsymbol{x}}

\newcommand*\diff{\mathop{}\!\mathrm{d}}

\def\be{\begin{equation}}
\def\ee{\end{equation}}
\def\bea{\begin{eqnarray}}
\def\eea{\end{eqnarray}}

\newcommand{\Photo}{}

\begin{document}

\begin{flushright}
\mbox{}
DESY-23-135 
\end{flushright}

\vspace*{4cm}
\title{Scalar dark matter production from the inflaton}

\author{M. Pierre}

\address{Deutsches Elektronen-Synchrotron DESY, Notkestr. 85, 22607 Hamburg, Germany}

\maketitle\abstracts{A curved space-time is known to act as a source for particle production in environments where gravity plays a significant role. We explore this effect in a minimal setup of cosmic inflation on the production of a scalar dark matter candidate during and after the inflationary stage of the universe. We consider the production of dark matter via direct coupling to the inflaton field or from pure gravitational interactions. Cosmological constraints from structure formation and dark matter isocurvature perturbations are discussed. A new analytical expression for the isocurvature power spectrum is provided.}

\section{Dark matter production during and after inflation}

We consider a minimal setup where a scalar dark matter (DM) candidate $\chi$ couples to a scalar inflaton field $\phi$  via a quartic coupling $\sigma$ whose general action reads
\begin{equation}
    \label{eq:action}
    \mathcal{S} \; = \; \int \diff^4 x \sqrt{-g} \left[-\frac{1}{2}M_P^2R + \frac{1}{2} (\partial_{\mu} \phi)^2 - V(\phi)  + \frac{1}{2} (\partial_{\mu} \chi)^2 - \frac{1}{2}m_{\chi}^2 \chi^2  - \dfrac{\sigma}{2} \phi^2 \chi^2\right]  \, .
\end{equation}
For concreteness, we consider the T-model~\cite{Kallosh:2013maa} inflationary potential $V(\phi)= 6 \lambda M_P^4 \tanh^2(\phi/\sqrt{6}M_P) $
that can be normalized for a nominal choice of $N_*=55$ $e$-folds for the fiducial CMB scale crossing $k_*=0.05~\text{Mpc}^{-1}$. The choice $\lambda \simeq 2 \times 10^{-11}$ corresponds to a spectral tilt $n_s\simeq 0.963$, tensor-to-scalar ratio $r \simeq 0.004$ and amplitude of scalar-perturbations power spectra $A_{S^*}$ compatible with current CMB measurements. Close to the minimum $\phi\ll M_P$, the potential can be approximated by $V(\phi)\simeq (m_\phi^2/2) \phi^2$ where the inflaton mass can be expressed as $m_{\phi} = \sqrt{2\lambda}M_P \simeq 1.6 \times 10^{13}~\text{GeV}$. Introducing the rescaled DM field $X \equiv a \chi$,
varying the action~(\ref{eq:action}) with respect to $X$ and moving to Fourier space, we obtain the equation of motion expressed in terms of the conformal time $\diff \eta = \diff t/a$ with $a$ being the scale factor and $t$ the cosmic time~\cite{Garcia:2022vwm}
\begin{equation}
    \label{eq:eom}
    (\partial_{\eta}^2 + \omega_k^2)X_k = 0,\quad \text{with} \quad \omega_k^2=k^2 + a^2 m_{\rm{eff}}^2 \quad \text{and} \quad m_{\rm{eff}}^2 = m_{\chi}^2 + \sigma \phi^2 + \frac{1}{6}R \, ,
\end{equation}
where the effective DM mass $m_{\rm{eff}}$ is time dependent. In this work we consider only scalars minimally coupled to gravity. A non-minimal DM coupling to gravity can manifest itself as an extra term in the effective mass squared. Extension of this discussion to non-minimal couplings can be found in~\cite{Garcia:2023qab}. \par \medskip During inflation, the quasi-homogeneous inflaton field $\phi(t)$ strongly affects the background and space-time curvature. The corresponding Ricci scalar $R$ causes the squared-mass to become negative and induces a tachyonic excitation of light scalar fields for wavelengths stretched on super-horizon scales, resulting in copious particle production. After the end of inflation $a>a_\text{end}$, the fast inflaton oscillations about the quadratic minimum are also responsible for particle production, induced via direct coupling $\sigma$ or by pure gravitational interactions, i.e. via $\lambda$, in the form of the Ricci scalar in Eq.~(\ref{eq:eom}).

\par \medskip

 Considering initially the absence of any DM particle corresponds to the Bunch-Davies vacuum initial condition, given by $\lim \limits_{\eta \rightarrow - \infty} X_k(\eta) = \frac{1}{\sqrt{2k}} e^{-i k \eta}$. The comoving number density of produced scalar DM, $n_{\chi}$, can be computed using the following expression~\cite{Kofman:1997yn,Ling:2021zlj}:
\begin{equation}
    \label{eq:comovingnd}
    n_{\chi} \left(\frac{a}{a_{\rm{end}}} \right)^3 \; = \; \int_{k_0}^{\infty} \diff k \frac{k^3}{2\pi^2}f_{\chi}(k, t) ~~{\rm{with}}~~ f_{\chi}(k, t) \, \equiv \, \frac{1}{2\omega_k}\left| \omega_k X_k - i X'_k \right|^2 \, ,
\end{equation}
with $f_{\chi}$ being the Phase Space Distribution (PSD). By defining the rescaled dimensionless comoving momentum $q=(k/m_\phi)(a/a_\text{end})$, one can distinguish modes with $q<1$ and $q>1$ that are respectively excited during or after inflation. \par
\begin{figure*}[!t]
\centering
    \includegraphics[width=0.5\textwidth]{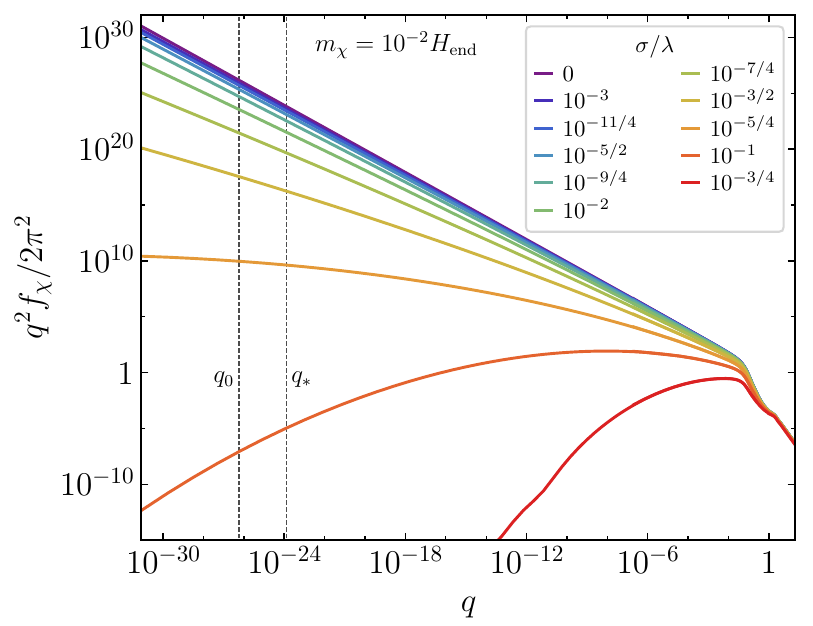}
 \hfill   \includegraphics[width=0.48\textwidth]{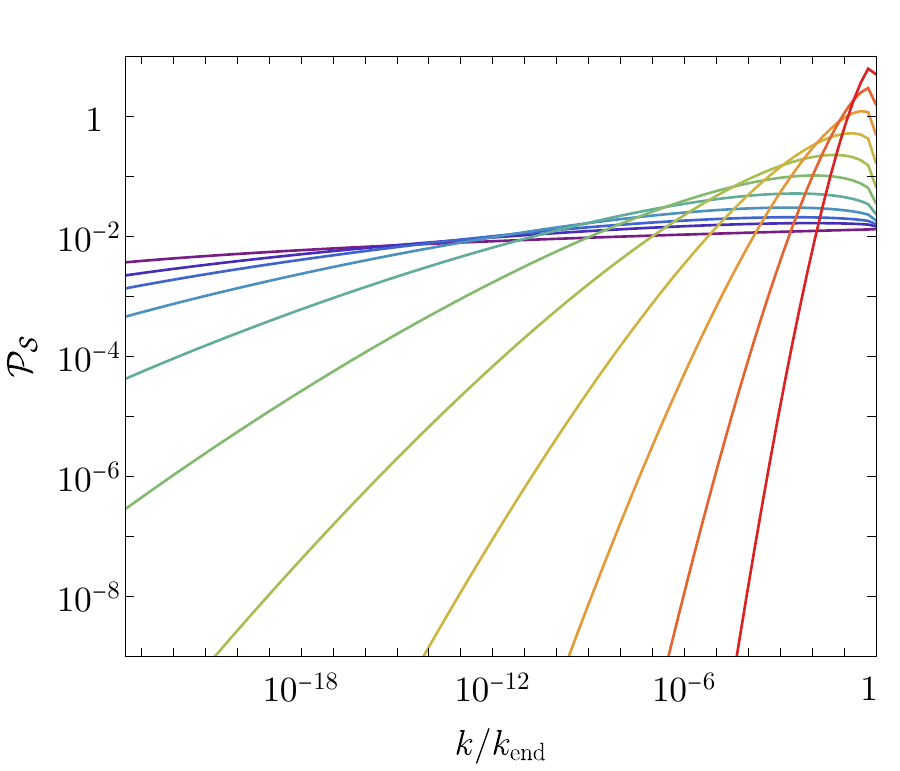}
    \caption{Dark matter phase space distribution (left) and isocurvature power spectrum (right) for selected values of coupling ratio $\sigma/\lambda$. The colour of each line codes the value of the coupling ratio $\sigma/\lambda$ and is identical on both left and right panels. }
    \label{fig:PSDs}
\end{figure*}

 
\noindent
\textbf{Modes excited during inflation.} For $q<1$, as discussed further on the phase space distribution scales as $f_\chi \sim q^{2 m_\text{eff}^2/(3H^2) - 3}$. If $\sigma=0$, the DM production is purely gravitational and the PSD scales as $f_{\chi} \propto q^{-3}$ for light  scalar DM $m_{\rm{\chi}} \ll H$. As $\sigma$ increases, the tachyonic excitation of the DM scalar field becomes suppressed, resulting in a blue-tilt of the spectrum at small $q$. A numerical evaluation of the PSD is represented on the left panel of Fig.~\ref{fig:PSDs}.

\noindent
\textbf{Modes excited after the end of inflation.} Such modes with $q>1$ never experience a tachyonic excitation as they remain sub-horizon during inflation. For mild values of $0<\sigma/\lambda<10^2$, both production via direct or gravitational couplings, i.e. induced via $\sigma$ or $\lambda$, result in a PSD scaling as $f_{\chi} \propto q^{-9/2}$ and increasing with $(\sigma/\lambda)^2$. For larger couplings $\sigma/\lambda>10^2$, strong parametric resonances induced by the oscillating mass term in Eq.~(\ref{eq:eom}) result in the formation of quasi-stochastic peaks in the PSD which has to be evaluated numerically. 

\par \medskip 

\section{Structure formation constraints}

Light DM candidates produced from a state out of thermal equilibrium may possess a sizable pressure component. Such pressure can suppress overdensities on galactic scales during the matter domination era, resulting in a cutoff in the matter power spectrum $\mathcal{P}(k)$ for scales $k$ larger than the free-streaming horizon wavenumber $
k_\text{H}(a) \, = \, \left( \int_0^a k_\text{FS}^{-1}(\tilde a) \diff \log \tilde a \right)^{-1}$, function of the free-streaming wave number $k_\text{FS}$ and the expansion history of the universe.  Measurements from the Lyman-$\alpha$ forest absorption-lines constrain this cutoff scale to $k_\text{H}(a=1)> 15 \, h \,\text{Mpc}^{-1}$. It can be translated into a bound on a generic warm dark matter (WDM) candidate initially thermally coupled to the SM bath $m_\text{WDM}>m_\text{WDM}^{\text{Ly}\mbox{-}\alpha} \; \sim \; 3~\text{keV}$. The free-streaming wave number $k_\text{FS} \, = \, \sqrt{9 \mathcal{H}^2/(10 w_\chi)} $ is uniquely a function of the conformal Hubble rate  $\mathcal{H}\equiv a H$ and the DM equation of state $w_\chi$ which can be approximated as
\begin{equation}
w_\chi \, \simeq \, \dfrac{T_\star^2}{3 m_\chi^2}\dfrac{\langle q^2 \rangle}{a^2} \,, \qquad \text{with} \qquad     \langle q^2\rangle \, \equiv \, \dfrac{\int \diff q\, q^4 f_\chi(q) }{\int \diff q\, q^2 f_\chi(q) } \,, \qquad \text{and} \qquad   T_\star \, \equiv \, m_\phi \left( \dfrac{a_\text{end}}{a_0} \right)	 \,,
\label{eq:EOSchi}
\end{equation}  
Since only the equation of state controls the cutoff scale in the matter power spectrum, one can translate the WDM bound to our scenario, by using the procedure introduced in~\cite{Ballesteros:2020adh}, via
\begin{equation}
 m_\chi\,>\,m_{\chi}^{\text{Ly}\mbox{-}\alpha} \;=\; m_{\rm WDM}^{\text{Ly}\mbox{-}\alpha} \left(\frac{T_{\star}}{T_{\rm WDM,0}}\right)\sqrt{\frac{\langle q^2\rangle}{\langle q^2\rangle_{\rm WDM}}}\,,
 \label{eq:lyalphaconst2}
\end{equation}
where $\langle q^2\rangle_{\rm WDM}\simeq 12.93$ $T_{\rm WDM,0}$ is the WDM temperature saturating the dark matter abundance.

\noindent
\textbf{Constraints for $\sigma/\lambda>1$.} In this case, the second moment $\langle q^2 \rangle$ is UV dominated and depends on the reheating temperature $T_\text{reh}$ through $\langle q^2 \rangle  \simeq  2.43  \sqrt{a_\text{reh}/a_\text{end}}$. This scaling allows to cancel out the $T_\text{reh}$ dependence on the dark matter energy-density at the present epoch, which asymptotes to $m_\chi \, > \, 32.4~\text{eV} $, almost independently of $\sigma/\lambda$ for $\sigma/\lambda>1$. However, parametric resonances induce minor deviations from this relation as represented numerically in Fig.~\ref{fig:bounds}.

\noindent
\textbf{Constraints for $\sigma/\lambda<1$}. In this case, the PSD and corresponding second moment become sensitive to the IR part. In that case, the constraint has to be evaluated numerically. As the coupling $\sigma/\lambda$ decreases, irrespectively of $\sigma/\lambda \ll 1$ the Lyman-$\alpha$ bound asymptotes to $m_\chi \, > 2 \times 10^{-4}~\text{eV}$. The full numerical evaluation of the constraints are depicted in Fig.~\ref{fig:bounds}, showing a good agreement with the analytical approximations provided in this section.

\section{Isocurvature perturbations}

The presence of a light scalar spectator field during inflation can lead to large isocurvature perturbations~\cite{Chung:2004nh}, in conflict with CMB measurements. The rapid increase in DM energy density is essentially driven by the quadratic fluctuations that substantially contribute to the variance $\langle \chi^2\rangle$~\cite{Chung:2004nh,Ling:2021zlj}. Initially assumed to vanish, the DM misalignment remains $\langle \chi\rangle = 0$ throughout inflation. The DM inhomogeneities do not directly affect the curvature perturbation, and they can be treated as pure isocurvature fluctuations in the comoving gauge~\cite{Chung:2004nh}. The second-order contribution to the isocurvature power spectrum is given by~\cite{Chung:2004nh,Ling:2021zlj} $\mathcal{P}_{\mathcal{S}}(k) \;=\; k^3/(2\pi^2\rho_{\chi}^2)\int \diff^3\bx \ \langle \delta\rho_{\chi}(\bx)\delta\rho_{\chi}(0) \rangle e^{-i \bk\cdot\bx}\,, $ where $\rho_{\chi}$ and $\delta\rho_{\chi}$ denote the DM energy density and its fluctuation, respectively. The current constraints on the isocurvature power spectrum from ${\textit Planck}$ impose $\mathcal{P}_{\mathcal{S}}(k_*)\lesssim 8.3 \times 10^{-11}$ for the pivot scale $k_*=0.05\,{\rm Mpc}^{-1}$. \par \medskip

\begin{figure*}[!t]
\centering
    \includegraphics[width=0.75\textwidth]{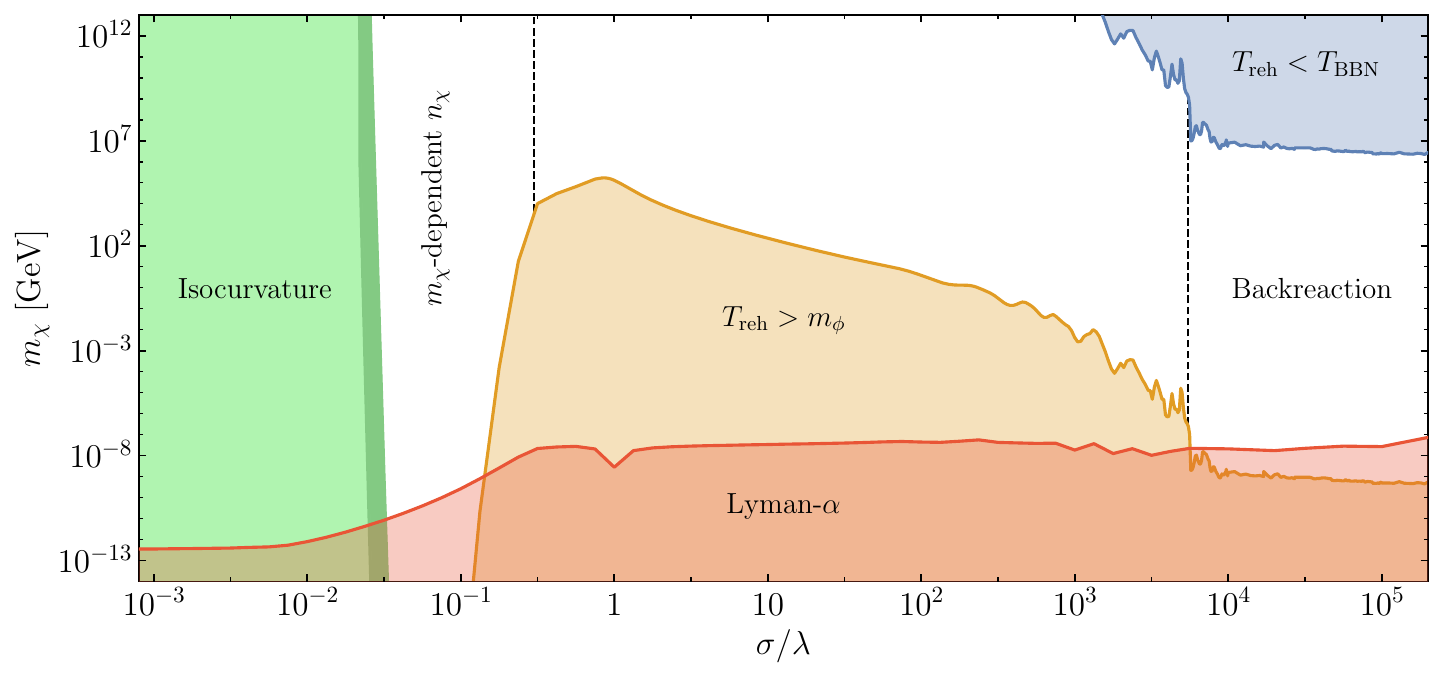}
    \caption{The white space represents the allowed parameter space for the DM mass and its coupling to the inflaton, in which the relic abundance constraint can be saturated. Forbidden regions correspond to the overproduction of isocurvature (green) and the suppression of small structure (red). Requiring the reheating temperature to be above that required for successful Big Bang Nucleosynthesis excludes the blue region. Ensuring that perturbative reheating can be achieved excludes the orange region.}
    \label{fig:bounds}
\end{figure*}

To derive an analytical expression for the isocurvature power spectrum, we consider for simplicity a finite duration of inflation $N_\text{tot}=76.5$ from $N_i=N_\text{end}-N_\text{tot}$ until $N_\text{end}$ which corresponds to IR and UV cutoffs $k_\text{IR}= a_i H \simeq 10 ^{-10} k_*$ and $k_\text{UV}= a_{\rm{end}} H \simeq 10^{24} k_*$. We further assume a constant Hubble rate $H$ and dark matter mass. By neglecting time derivatives of the mode function, the isocurvature power spectrum can be expressed as~\cite{Garcia:2023awt}
\begin{equation}
\mathcal{P}_{\mathcal{S}}(k) \;=\; \frac{k^2}{(2\pi)^4\langle \chi^2 \rangle^2}\int_0^{\infty} \diff  p\ p \int_{|k-p|}^{k+p}\diff q\ q \ |\chi_p|^2 |\chi_q|^2\,, \qquad \text{with} \qquad     \langle \chi^2 \rangle \, \equiv \, \int_{k_\text{IR}}^{k_\text{UV}} \, \mathcal{P}_\chi \, \dfrac{\diff k}{k} \, ,
\label{eq:isocurvatureanalytical}
\end{equation}
with $    \mathcal{P_\chi} \, \equiv \, k^3/(2 \pi^2)| \chi_k |^2 $. Assuming $m_\chi^2/(3 H^2)<1$ , the mode function on super-horizon scales can be approximated by $   | \chi_k(k \ll aH) | \, \simeq \,    (2^{1/3} a H^{1/3})^{-3/2} (k/(aH))^{-\nu } \,,
$ with $\nu \simeq 3/2 - \beta$ and $\beta\equiv m_\chi^2/(3 H^2)$. The corresponding power spectrum $\mathcal{P_\chi} \,= H^2/(4 \pi^2) (k/(aH))^{ 2 \beta}$ is scale invariant for $\beta=0$ and blue-tilted when $\beta>0$.  Defining $ \Delta \hat N\equiv N_\text{end}-\hat N$ with $\hat N$ being the number of $e$-folds at horizon crossing for a given scale $ k=a(\hat N)H(\hat N)$, we can generalize the formula derived in~\cite{Garcia:2023awt} to arbitrary scales $k$ and time-dependent DM mass
\begin{equation}
     \mathcal{P}_{\mathcal{S}}(k) \;\simeq \; \dfrac{4 \sigma ^2 \hat \phi ^4 (N_\text{tot}- \Delta \hat N) }{9 \hat H^4 }  \, e^{-4 \Delta \hat N \sigma  \hat \phi ^2/(3 \hat H^2)} \, \left(1-e^{-2 N_\textrm{tot} \sigma \hat \phi ^2/(3 \hat H^2)}\right)^{-2} \,,
     \label{eq:approximatedisocurvature}
\end{equation}
where the $\,\hat{ }\,$ notation means that background quantities have to be evaluated at horizon crossing. Generalization to any inflaton-induced DM mass $m_\text{eff}(\phi)$ can be done by substituting $\sigma \hat \phi^2 \rightarrow m_\text{eff}^2(\hat \phi) $. From Eq.~(\ref{eq:approximatedisocurvature}), the isocurvature constraint $\mathcal{P}_{\mathcal{S}}(k_*) < \beta_\text{iso} \mathcal{P}_{\mathcal{R}}(k_*) \simeq  10^{-11}$ corresponds to $\sigma/\lambda \, > \,  0.02$ which is in good agreement with the fully numerical result shown in Fig.~2 left-panel of Ref.~\cite{Garcia:2023awt} and depicted in Fig.~\ref{fig:bounds}. The available parameter space is represented as a white region in Fig.~\ref{fig:bounds} while coloured region are excluded. See Ref.~\cite{Garcia:2022vwm} for details.

\section*{Acknowledgments}

I acknowledge support by the Deutsche Forschungsgemeinschaft (DFG, German Research Foundation) under Germany's Excellence Strategy – EXC 2121 “Quantum Universe” – 390833306.

\end{document}